\documentclass[utf8]{frontiersSCNS} 

\usepackage{hyperref}
\hypersetup{colorlinks,linkcolor={blue},citecolor={blue},urlcolor={red}}  

\usepackage[onehalfspacing]{setspace}

\usepackage{natbib}

\DeclareUnicodeCharacter{22C6}{*}

\def\keyFont{\fontsize{8}{11}\helveticabold }
\def\firstAuthorLast{Nsamba {et al.}} 
\def\Authors{Benard Nsamba\,$^{1,2,*}$, Tiago L.~Campante\,$^{1,2}$, Mário J.~P.~F.~G.~Monteiro\,$^{1,2}$, Margarida S.~Cunha\,$^{1,2}$ and Sérgio G.~Sousa\,$^{1,2}$}
\def\Address{$^{1}$Instituto de Astrof\'{\i}sica e Ci\^{e}ncias do Espa\c{c}o, Universidade do Porto,  Rua das Estrelas, PT4150-762 Porto, Portugal\\
$^{2}$Departamento de F\'{\i}sica e Astronomia, Faculdade de Ci\^{e}ncias da Universidade do Porto, PT4169-007 Porto, Portugal
}
\def\corrAuthor{Benard Nsamba}

\def\corrEmail{benard.nsamba@astro.up.pt}

\begin{document}
\onecolumn
\firstpage{1}

\title[Impact of the metallicity mixture]{On the nature of the core of $\alpha$ Centauri A: the impact of the metallicity mixture

} 

\author[\firstAuthorLast ]{\Authors} 
\address{} 
\correspondance{} 

\extraAuth{}

\maketitle

\begin{abstract}
Forward asteroseismic modelling plays an important role towards a complete understanding of the physics taking place in deep stellar interiors. With a dynamical mass in the range over which models develop convective cores while in the main sequence, the solar-like oscillator $\alpha$ Centauri A presents itself as an interesting case study. We address the impact of varying the metallicity mixture on the determination of the energy transport process at work in the core of $\alpha$ Centauri A. We find that $\gtrsim$ 70$\%$ of models reproducing the revised dynamical mass of $\alpha$ Centauri A have convective cores, regardless of the metallicity mixture adopted. This is consistent with the findings of \citeauthor{201Nsamba}, where nuclear reaction rates were varied instead. Given these results, we propose that $\alpha$ Centauri A be adopted in the calibration of stellar model parameters when modelling solar-like stars with convective cores.

\tiny
 \keyFont{ \section{Keywords:} HD 128620, Asteroseismology, Stellar modelling, Fundamental parameters, Convection, Radiation
} 
\end{abstract}

\section{Introduction}

Most parameters used in stellar modelling are calibrated based on the Sun, e.g., the mixing length parameter, the helium-to-heavy-element ratio, chemical abundances etc. This approach is well-suited to the modelling of stars with similar properties to the Sun, i.e., for solar-type stars with a mass below 1.1 M$_\odot$. The quest for a more massive star with well-known properties and interior structure is of the utmost importance, as such star could become a potential model calibrator for solar-like stars having convection as the main energy transport process in their cores.

The bright, nearby binary $\alpha$ Centauri is amongst the best characterized star systems, with a plethora of available high-precision observations, e.g., parallaxes, angular diameters, interferometric radii, metallicities, effective temperatures, luminosities, and oscillation frequencies (\citealt{derhjelm,Pour2002,Bouchy,Bed,Bazot2007,Meulen2010,Pour2016,Kervella2016,Kervella2017}). $\alpha$ Centauri A is of particular interest, since its dynamical mass is in the range (1.1 -- 1.15 M$_\odot$; \citealt{Aerts2010}) over which models constructed at solar metallicity are expected to develop convective cores while in the main sequence. This has given rise to studies that aimed at establishing the nature of its core and at exploring the physics that affect core properties.
 
Forward stellar modelling, when coupled with observational constraints from asteroseismology, constitutes a valuable tool in the exploration of the physics of stellar interiors (e.g., \citealt{Monta,Mathur,Still,Metcalfe,Lebreton,Aguirre,2018Nsamba}). \citet{2016Bazot} complemented asteroseismic and spectroscopic observables with the dynamical mass determined by \citet{Pour2002} to tightly constrain stellar models of $\alpha$ Centauri A. They explored the impact of varying the nuclear reaction rates, overshoot, and diffusion of hydrogen and helium, having found a noticeable change in the number of best-fit models\footnote{In this work, models that are representative of a set of observables are termed as best-fit models.} with convective cores when nuclear reaction rates were varied. Following the revision of the dynamical mass of $\alpha$ Centauri A (\citealt{Pour2016}), \citet{201Nsamba} (hereafter Paper I) carried out a detailed modelling of this star again allowing the nuclear reaction rates to vary, and found about 70\% of best-fit models to have convective cores. More recently, \citet{2018Joyce} suggested that, if $\alpha$ Centauri A has a convective core, then it would be necessary to modify standard physical prescriptions (e.g., enhancing diffusion) in order to correctly model the star. Amongst the different model physics explored in Paper I and in \citet{2016Bazot}, the impact of the metallicity mixture on the core properties of $\alpha$ Centauri A has, however, not been investigated.
 
In this work we investigate the impact of the metallicity mixture on the inferred nature of the core of $\alpha$ Centauri A. The paper is organized as follows: Section \ref{model_grids} describes the stellar model grids and sets of observables used in the optimization process. A discussion of the results is presented in Sect.~\ref{discuss}. Section~\ref{conclude} presents the conclusions.

\section{Model grids and observational constraints}
\label{model_grids}

To explore the impact of varying the metallicity mixture on the nature of the core of $\alpha$ Centauri A, we set up two grids (A and B) with the same model physics except for the metallicity mixture (see Table \ref{mixtures}). The stellar evolution code MESA (Modules for Experiments in Stellar Astrophysics; \citealt{Pax1,Pax2,Pax3,Pax2018}) version 9793 was used to generate the grids. 

We set the metallicity mixture in Grid A according to \citet{Grevesse} with a solar surface heavy element mass fraction $Z_{\rm surface,\odot}=0.016$, while Grid B uses the metallicity mixture from \citet{Asplund} with $Z_{\rm surface,\odot}=0.0134$. The main motivation for considering these two mixtures goes back to the theoretically predicted sound speed profiles for solar models constructed with the different composition mixtures. \citet{Delahaye2006}
reported that solar models using the \citet{Grevesse} and \citet{Asplund} mixtures yield a sound speed profile close to that of the real Sun, as opposed to models that use the \citet{Asplund2005} mixture.
We recall that the metallicity, [Fe/H], is defined as:
\begin{equation}
\mbox{[Fe/H]} = \log\left( \frac{Z_{\rm surface}}{X_{\rm surface}}\right)_{\rm star} - ~\log\left( \frac{Z_{\rm surface}}{X_{\rm surface}}\right)_\odot \, ,
\end{equation}
where $X_{\rm surface}$ is the surface hydrogen mass fraction. Notice that varying the metallicity mixture requires setting the corresponding appropriate opacities. We used opacities from OPAL tables (\citealt{Iglesias}) at high temperatures, whereas at low temperatures tables from \citet{Ferguson} were used instead, for the respective metallicity mixtures. We employed the Joint Institute for Nuclear Astrophysics Reaction Library (JINA REACLIB; \citealt{Cyburt}) in both grids. The $^{14}{\rm N}(p,\gamma)^{15}{\rm O}$ and $^{12}{\rm C}(\alpha, \gamma)^{16}{\rm O}$ rates were described according to \cite{Imbriani} and \cite{Kunz}, respectively. We note that the $^{12}{\rm C}(\alpha, \gamma)^{16}{\rm O}$ reaction rate is less relevant for stars on the main-sequence phase but is vital in more evolved stars, i.e., stars at the core helium-burning evolution stage.
\begin{table}[!h]
\centering 
\caption{Main features of the model grids adopted in this work.}
\begin{tabular}{cccc}        %
\hline
Grid	              & Metallicity Mixture	&  Core Overshoot	&	Diffusion	       \\
\hline
A	                  & \citet{Grevesse}	   	&  Yes				&   Yes							\\
B  	  		          & \citet{Asplund}  	 	&  Yes  			&   Yes							\\
\hline                                   
\end{tabular}
\label{mixtures}
\end{table}
Table \ref{mixtures} lists the macrophysics used in either grid. We note that core overshoot was included as described by \citet{Herwig} for models identified to have developed convective cores. Atomic diffusion was included in all our models according to \citet{Thoul}. The latter is known to be a vital chemical transport process in low-mass stars (e.g., \citealt{Aguirre1,2018Nsamba}). The mixing length theory, as described by \citet{Vitense}, was used to describe convection. We also implemented the Grey--Eddington atmosphere to integrate the atmospheric structure from the photosphere to an optical depth of $10^{-4}$. The helium mass fraction ($Y$) was estimated using the relation
\begin{equation}
    Y = Z\left(\frac{\Delta Y}{\Delta Z}\right) + Y_{0} \, ,
\end{equation}
where $\Delta Y / \Delta Z $ is the helium-to-heavy-element relation (set to 2.0; \citealt{Serenelli}) and $Y_0$ is the big bang nucleosynthesis value (set to 0.2484; \citealt{Cyburt2003}).

Evolutionary tracks were evolved from the zero-age main sequence to the end of the subgiant evolution phase. The terminal criterion affecting the tracks is similar to that implemented in Paper I. The parameter space of the model grids is as follows: $M \in$ [1.0, 1.2] $\rm M_\odot$ in steps of 0.01 $\rm M_\odot$; mixing length parameter, $\alpha_{\rm mlt}$ $\in$ [1.3, 2.5] in steps of 0.1; overshoot parameter, $f_{\rm ov}$ $\in$ [0, 0.03] in steps of 0.005; and $Z \in$ [0.023, 0.039] in steps of 0.005. Each model grid contains about 156,000 models. The corresponding adiabatic oscillation frequencies of each model, for spherical degrees $l$ = 0, 1, 2, and 3, were determined using GYRE \citep{Townsend}. The surface effects were accounted for using the combined-term surface correction method described by \citet{Ball2014}.
This surface correction method has been reported to yield the least internal systematic uncertainties among the different available corrections (e.g., \citealt{2018Nsamba,Compton2018,gensen2019}).

Table \ref{adopted} displays the spectroscopic and interferometric constraints used in the optimization process. Run I adopts $T_{\rm eff}$ and [Fe/H] values obtained in Paper I.
 \begin{table}[!h]
\centering 
\caption{Spectroscopic and interferometric constraints.}
\begin{tabular}{cccc}        %
\hline 
Run	             & $T_{\rm eff}$ (K)		&  [Fe/H] (dex)		&	Radius	($\rm R_\odot$)       \\
\hline
I	             & 5832 $\pm$ 62   	   		&  0.23 $\pm$ 0.05	&  1.2310 $\pm$ 0.0036	\\
II  	  		     & 5795 $\pm$ 19   	        &  0.23 $\pm$ 0.05  &  1.2234 $\pm$ 0.0053	\\
\hline                                   
\end{tabular}
\label{adopted}
\end{table}
 These spectroscopic constraints were complemented with the interferometric radius from \citet{Pour2016}. Run II adopts $T_{\rm eff}$ and interferometric radius values from \citet{Kervella2016}. We further note that \citet{Pour2002} derived a dynamical mass of 1.105 $\pm$ 0.0070 M$_\odot$. This dynamical mass was then revised by \citet{Pour2016}, who obtained a value of 1.133 $\pm$ 0.0050 M$_\odot$. 

Finally, the same asteroseismic constraints as in Paper I were adopted. The Bayesian code AIMS (Asteroseismic Inference on a Massive Scale; \citealt{Reese,Rend2019}), a software for fitting stellar pulsation data, was used as our optimization tool. Stellar parameters and their associated uncertainties were taken as the mean and standard deviation of the resulting posterior probability distribution functions (PDFs), as output by AIMS.

\section{Discussion}
\label{discuss}

\begin{figure*}[!h]
\minipage{0.38\textwidth}
\hspace{1cm}
  \includegraphics[width=\linewidth]{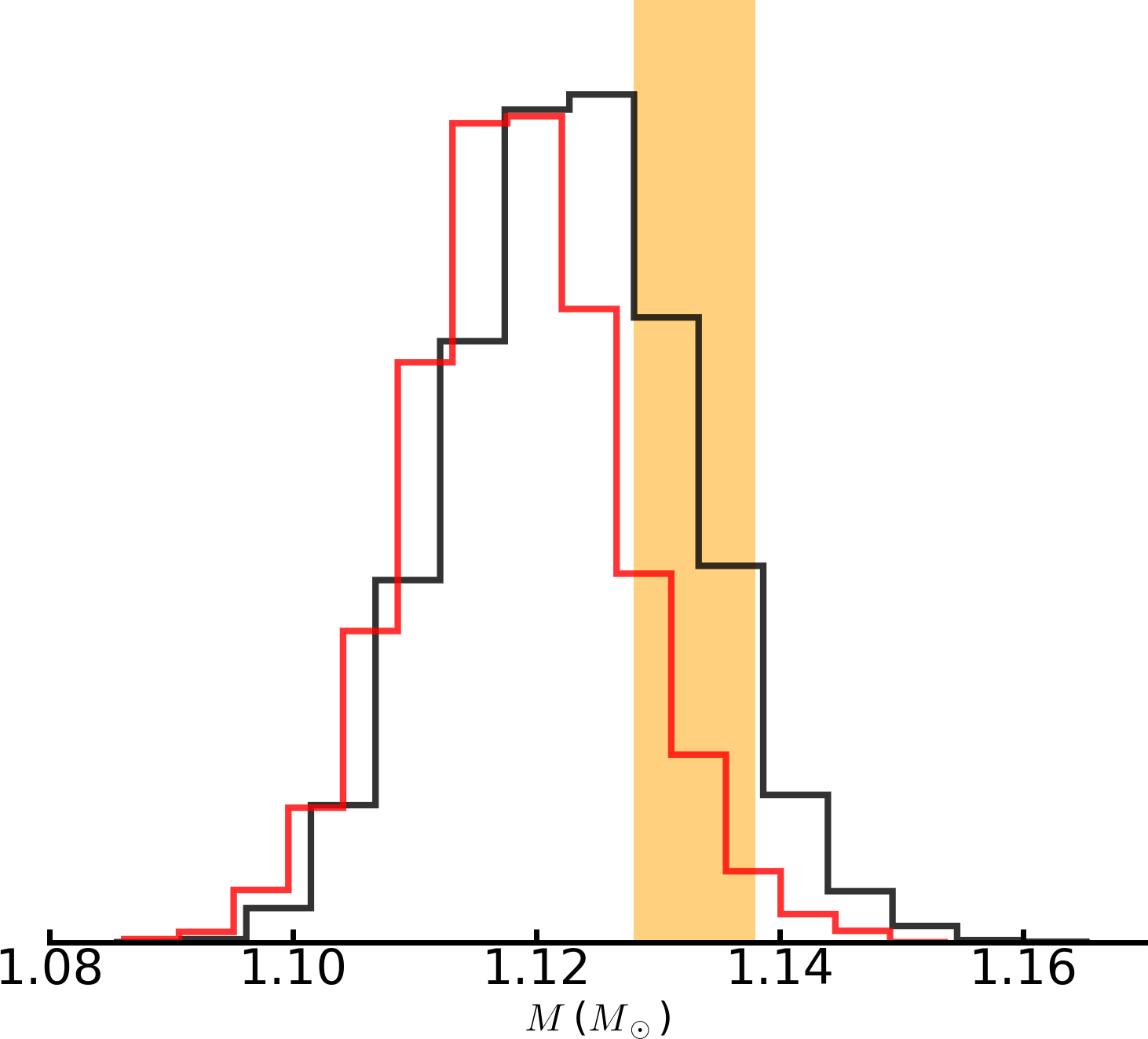}
\endminipage\hfill
\minipage{0.38\textwidth}
\hspace{-2cm}
  \includegraphics[width=\linewidth]{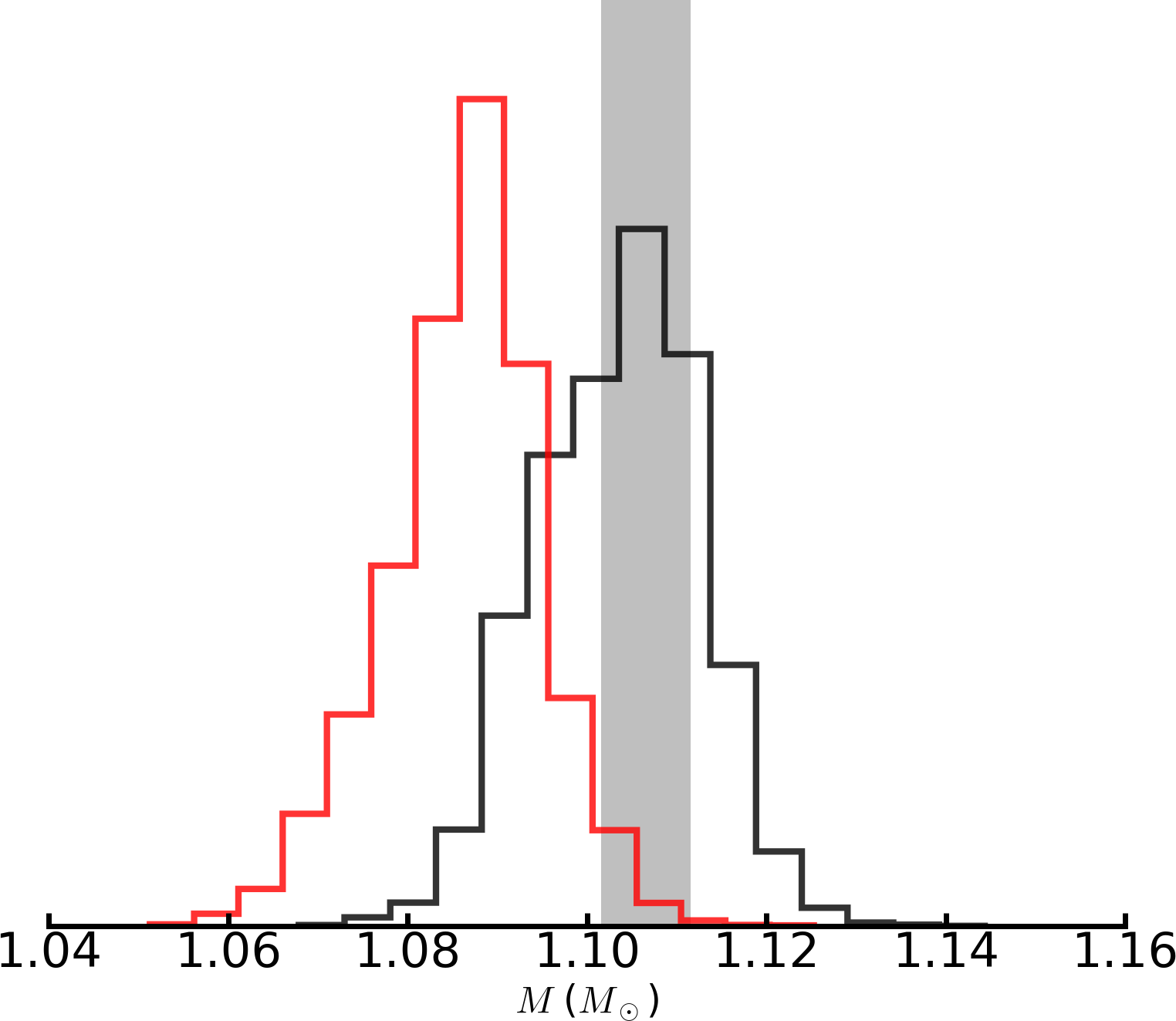}
\endminipage
\caption{Run I (left) and Run II (right): Histograms represent the stellar mass posterior PDFs obtained using Grids A (red) and B (black). The dynamical masses (and corresponding uncertainties) of \citet{Pour2016} and 
\citet{Pour2002} are shown in orange and grey, respectively.
}
\label{mass_parameter}
\end{figure*}

Table \ref{results} presents the stellar parameters determined using the model grids described in Table \ref{mixtures} and the sets of observables in Table \ref{adopted}. Our results show that we are able to reproduce the dynamical masses of \citet{Pour2016} (Run I) and \cite{Pour2002} (Run II) within 1$\sigma$ (see Fig.~\ref{mass_parameter} and Table \ref{results}). We note that the observed luminosity (i.e., 1.521 $\pm$ 0.015 $\rm L_\odot$; \citealt{Kervella2017}) of $\alpha$ Centauri A was not included among the sets of observables as shown in Table~\ref{adopted}. This is constrained via the combination of the interferometric radius and effective temperature. Our derived luminosity values are in agreement with the observed values (see Table~\ref{surf1}). Run II and Run I luminosity values from both grids are consistent within 1$\sigma$  and 2$\sigma$, respectively. The slight increase in luminosity values  obtained in Run I is attributed to the larger interferometric radius used (see Table~\ref{adopted}). 
\begin{figure*}[!h]
\minipage{0.36\textwidth}
\hspace{1cm}
  \includegraphics[width=\linewidth]{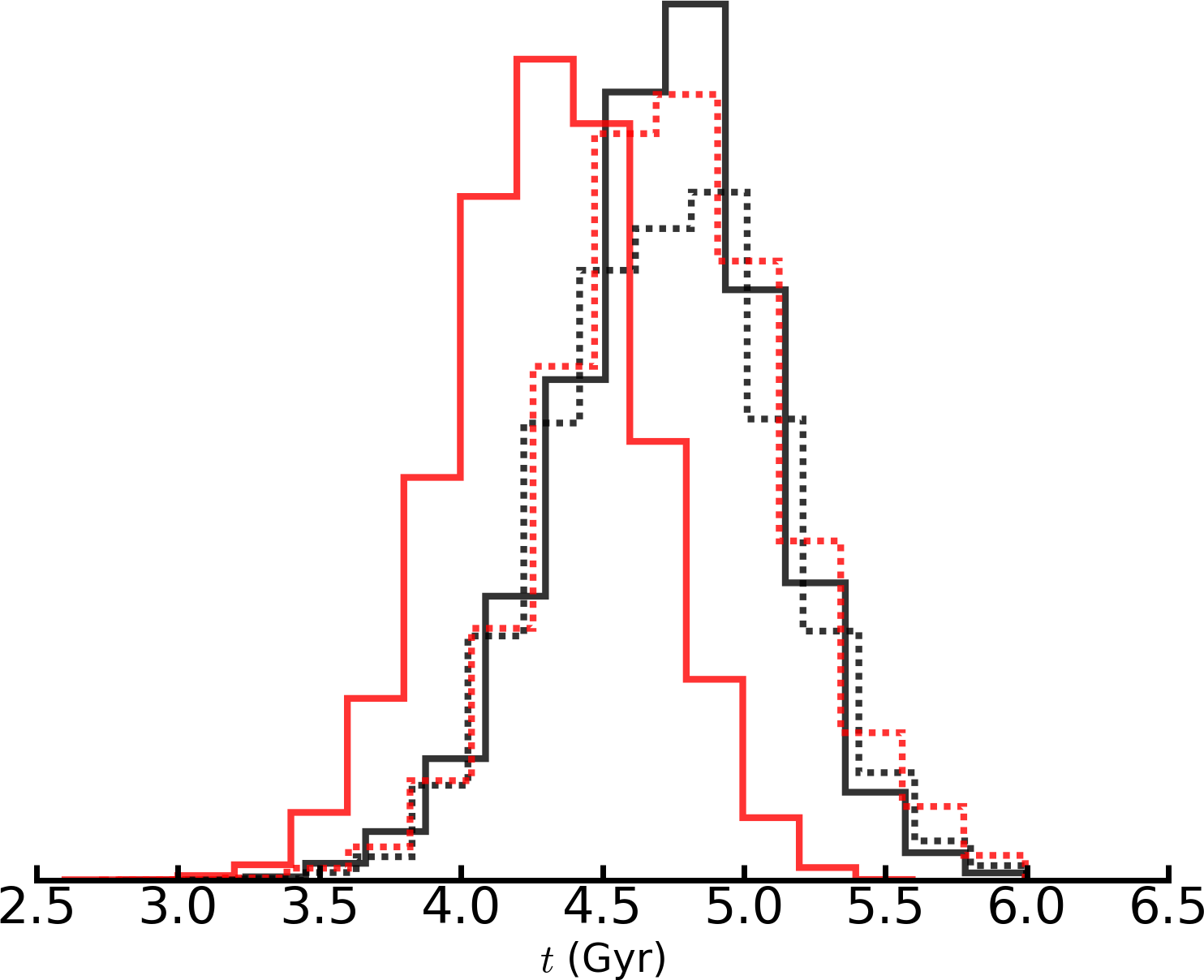}
\endminipage\hfill
\minipage{0.36\textwidth}
\hspace{-2cm}
  \includegraphics[width=\linewidth]{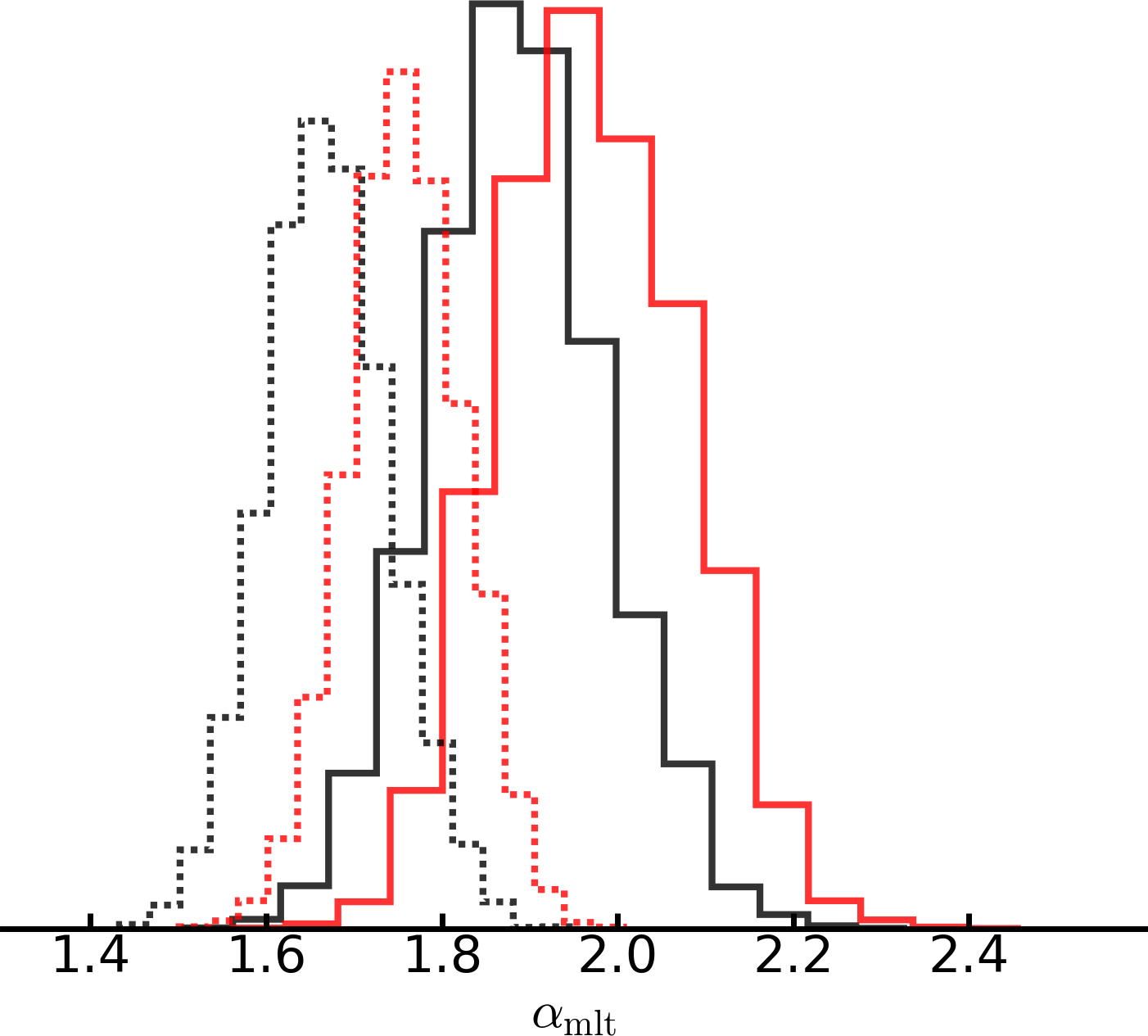}
\endminipage\hfill
\minipage{0.36\textwidth}%
\hspace{1cm}
  \includegraphics[width=\linewidth]{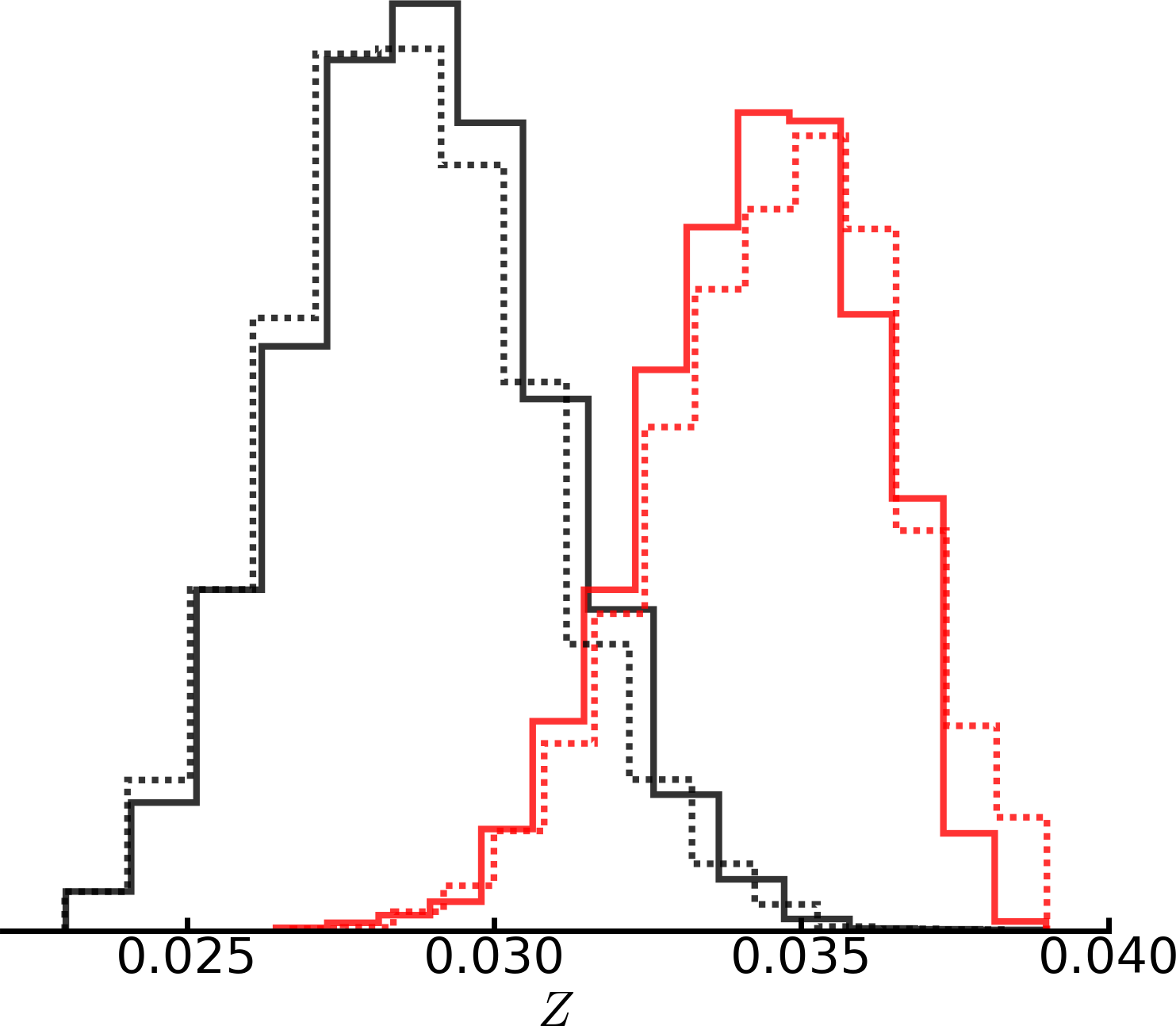}
\endminipage\hfill
\minipage{0.36\textwidth}
\hspace{-2cm}
  \includegraphics[width=\linewidth]{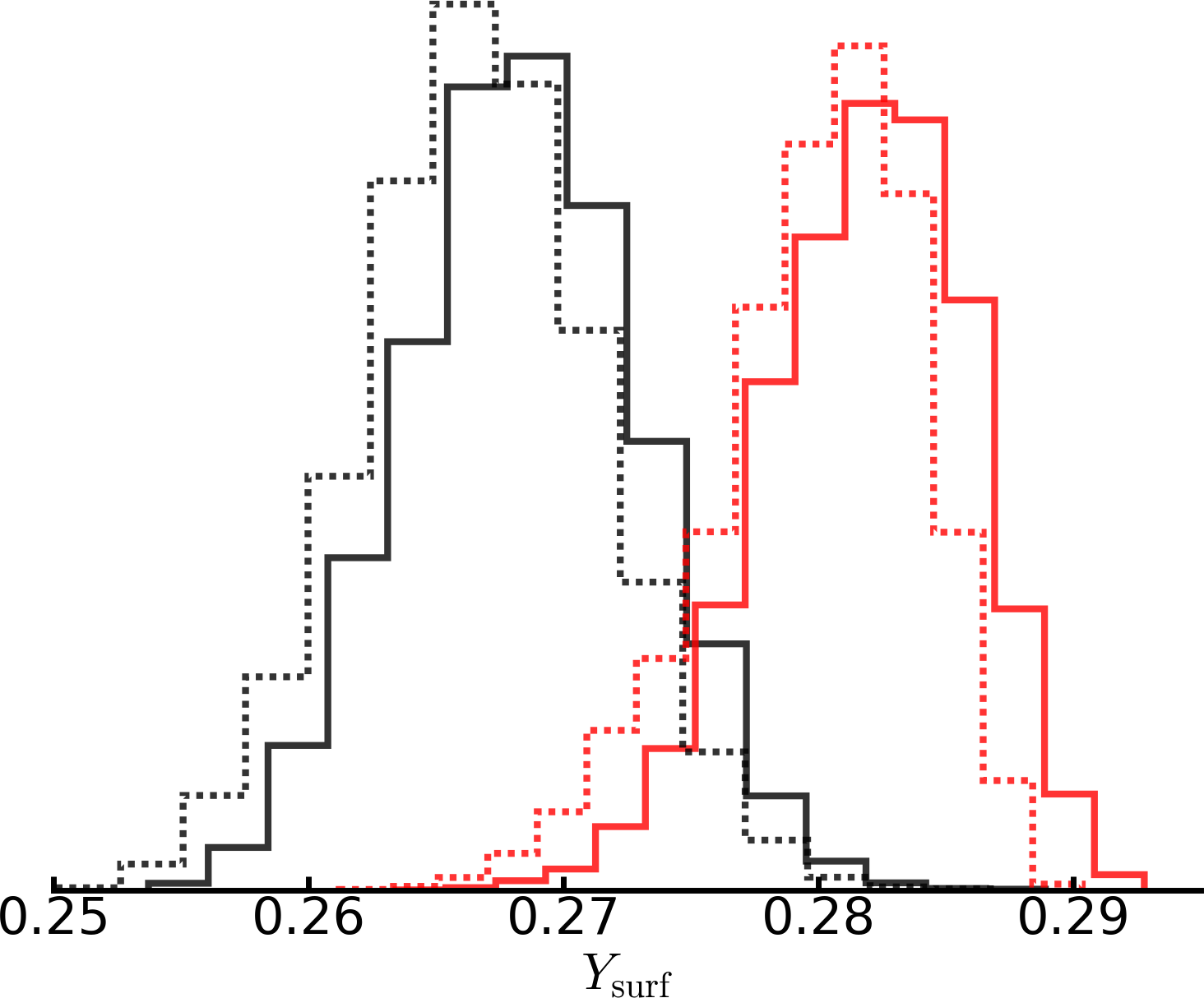}
\endminipage\hfill
\minipage{0.36\textwidth}%
\hspace{1cm}
  \includegraphics[width=\linewidth]{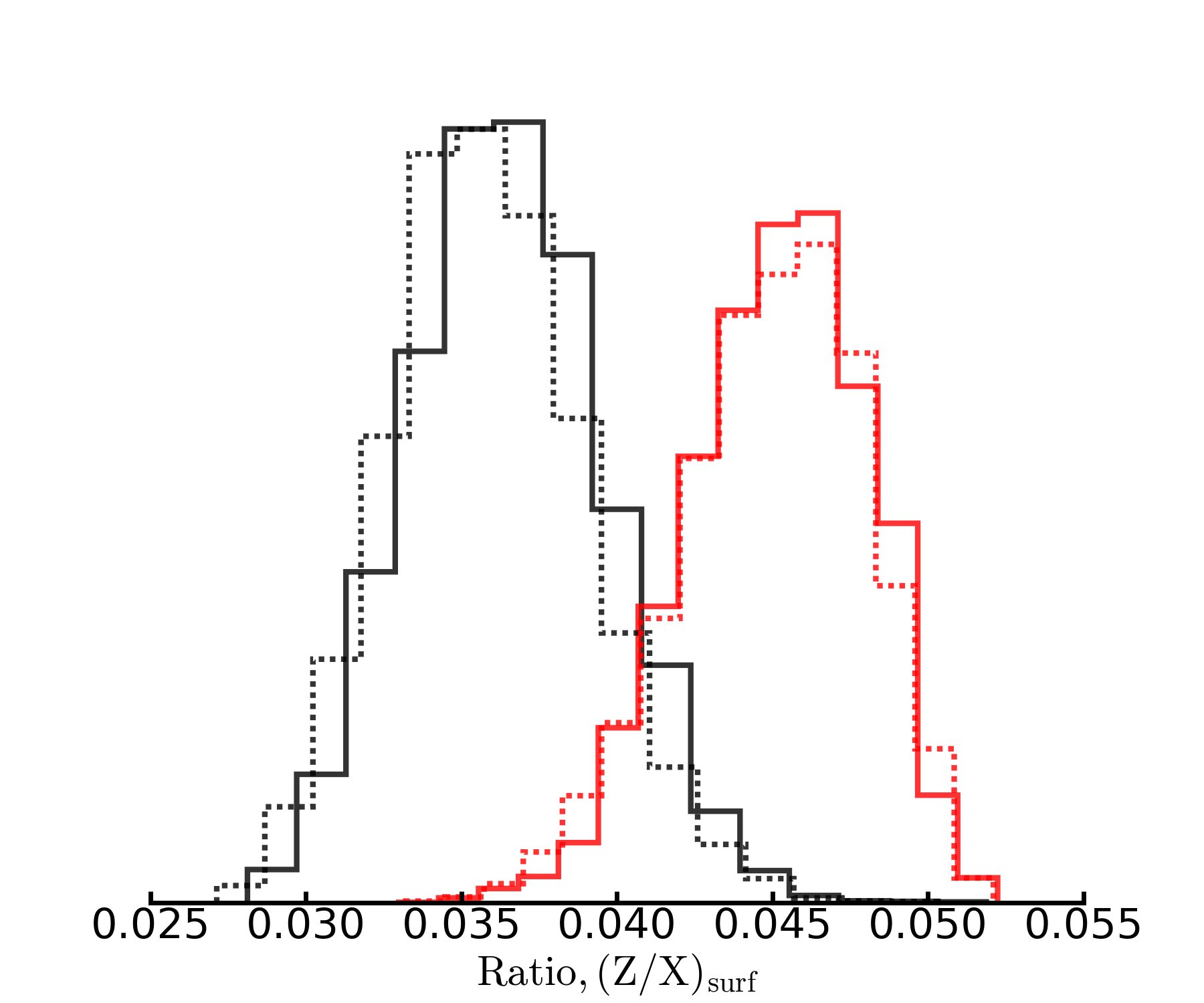}
\endminipage\hfill
\minipage{0.36\textwidth}%
\hspace{-2cm}
  \includegraphics[width=\linewidth]{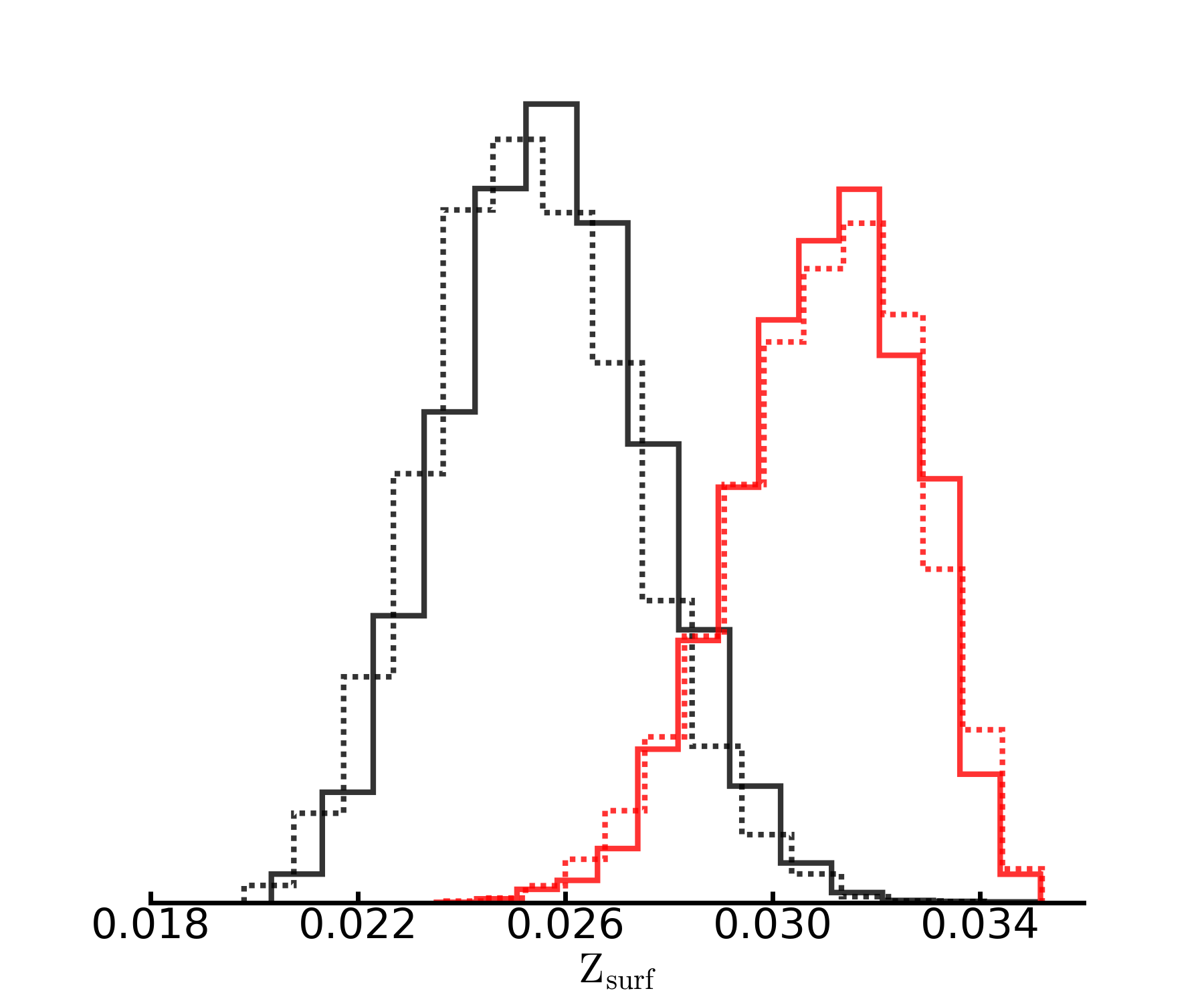}
\endminipage
\caption{Histograms represent stellar parameter posterior PDFs obtained using different grids (Grid A in red and Grid B in black) and observational constraints (Run I as solid lines and Run II as dashed lines).
}
 \label{other_parameters}
\end{figure*}

\begin{table*}[!h]
\centering 
\caption{Stellar parameters determined using different grids and observational constraints. The sixth column shows the percentage of best-fit models with convective cores, while the seventh column shows the upper convective-core mass limit.}
\begin{tabular}{ccccccc}        
\hline 
Grid	& Run 	&	$M$ $(\rm M_\odot)$		&  $t$ (Gyr) & $\alpha_{\rm mlt}$  & Convective Core (\%) & $M_{\rm c}$ ($\rm M_{\odot}$)\\
\hline

 A 		& I 		& 	1.12 $\pm$ 0.01		&  4.30 $\pm$ 0.35 	& 1.97 $\pm$ 0.10     & 70 &  $\leqslant$ 0.085 	   \\[0.2cm]
        & II		&	1.09 $\pm$ 0.01		&  4.74 $\pm$ 0.40 	 &  1.76 $\pm$ 0.07    & 46 & $\leqslant$ 0.084\\	
\hline 
 B 		& I 		&	1.12 $\pm$	0.01	&   4.72 $\pm$ 0.37	 &	1.89 $\pm$ 0.10   & 79 & $\leqslant$ 0.096\\[0.2cm]
        & II		&	1.10 $\pm$ 0.01		&	4.73 $\pm$ 0.39  & 1.67 $\pm$ 0.07      & 79 & $\leqslant$ 0.092 	\\
\hline 
\end{tabular}
\label{results}
\end{table*}
\begin{table*}[!h]
\centering 
\caption{Luminosities and abundances determined using different grids and observational constraints.}
\begin{tabular}{ccccccc}        
\hline 
Grid	& Run 	   &  $L$ (\rm L$_\odot$)	& $Z$ & $Y_{\rm surf}$ 	&  $Z_{\rm surf}$        	& $(Z/X)_{\rm surf}$    \\
\hline

 A 		& I 		& 1.703 $\pm$ 0.059 & 0.034 $\pm$ 0.002		& 0.282 $\pm$ 0.004	& 0.031 $\pm$ 0.002 &  0.045 $\pm$ 0.003  		\\
        & II		& 1.502 $\pm$ 0.023	&  0.035 $\pm$ 0.002	& 0.279 $\pm$ 0.004	& 0.031 $\pm$ 0.002  & 0.045 $\pm$ 0.003	\\
\hline 
 B 		& I 		& 1.675 $\pm$ 0.063	&  0.028 $\pm$ 0.002	& 0.269 $\pm$ 0.005 & 0.026 $\pm$ 0.002 & 0.036 $\pm$ 0.003   	\\
        & II		& 1.510 $\pm$ 0.022	&  0.028 $\pm$ 0.002	& 0.267 $\pm$ 0.005	& 0.025 $\pm$ 0.002 & 0.036 $\pm$ 0.003	 \\
\hline 
\end{tabular}
\label{surf1}
\end{table*}


A clear difference can be seen in the right panel of Fig.~\ref{mass_parameter} (Run II) between the stellar mass posterior PDFs obtained using the two grids. Since varying the metallicity mixture has been shown to have a minimum effect on the estimated stellar mass (\citealt{Aguirre,2018Nsamba}), this feature can instead be explained by the different core properties of the best-fit models. Table \ref{results} shows that the number of best-fit models with convective cores changes from 46\% to 79\% when the metallicity mixture is changed from that of \citet{Grevesse} to that of \citet{Asplund}. This happens since the dynamical mass of \citet{Pour2002} lies within a range in which the onset of the CNO (carbon–nitrogen–oxygen) cycle, and thus core convection, is highly sensitive to the adopted metallicity mixture.

A different scenario is found when considering models that reproduce the dynamical mass of \citet{Pour2016}, with the stellar mass posterior PDFs showing excellent agreement (Run I; see left panel of Fig.~\ref{mass_parameter}). The percentage of models with convective cores is now consistent (i.e., $\gtrsim$ 70$\%$) irrespective of the model grid adopted. We note that best-fit models are on average higher in mass compared to Run II and most have already developed convective cores, with any variation in the metallicity mixture generating no significant difference on their core properties.

In Table~\ref{results}, we show the upper limits of the convective-core mass ($M_{\rm c}$) of our best-fit models. From all runs, we find the core radius to have an upper limit of 0.11 $\rm R_\odot$. \citet{2012Bazot} derived an upper limit for the radius and the mass of a possible convective core in $\alpha$ Centauri A to be 0.059 $\rm R_\odot$ and 0.035 $\rm M_{\odot}$, respectively. These limits were derived while taking into account the small frequency separation ($\delta \nu$) in the optimisation, as this parameter can provide a direct estimation of the convective core characteristics. Furthermore, when exploring the contribution of the different model physics to the nature of the core of $\alpha$ Centauri A, \citet{2016Bazot}, in their table 3, report the core radius of their best-fit models to vary between 0.026 and 0.084 $\rm R_\odot$, which is consistent with our findings.

The top left panel of Fig.~\ref{other_parameters} shows that the derived stellar ages are in excellent agreement irrespective of the grid and observational constraints used. Furthermore, these ages are consistent with literature values  (\citealt{1999Kim,Yild,2016Bazot,201Nsamba,2018Joyce}). Table \ref{results} and the top right panel of Fig.~\ref{other_parameters} show that the $\alpha_{\rm mlt}$ estimated based on either Run I (solid lines) or Run II (dashed lines) are consistent within 1$\sigma$. The values of $\alpha_{\rm mlt}$ across runs are however different, this being mainly due to the different radius constraints used (see Table \ref{adopted}). We note that the interferometric radius measurements used in each run indirectly constrain the model mass. Moreover, $\alpha_{\rm mlt}$ is known to have a significant degree of correlation with the stellar mass and effective temperature \citep{Pinheiro}.

A clear contrast can be seen in the bottom panels of Fig.~\ref{other_parameters} between the best-fit models obtained using the two grids. As expected, the grid based on the metallicity mixture from \citet{Asplund} (Grid B) leads to best-fit models with a lower $Z$ compared to those based on the mixture from \citet{Grevesse} (Grid A). A similar feature can be seen for the surface helium mass fraction, $Y_{\rm surf}$. 

\begin{figure*}[!h]
\vspace{-1.6cm}
\centering
  \includegraphics[width=10cm, height=8.5cm]{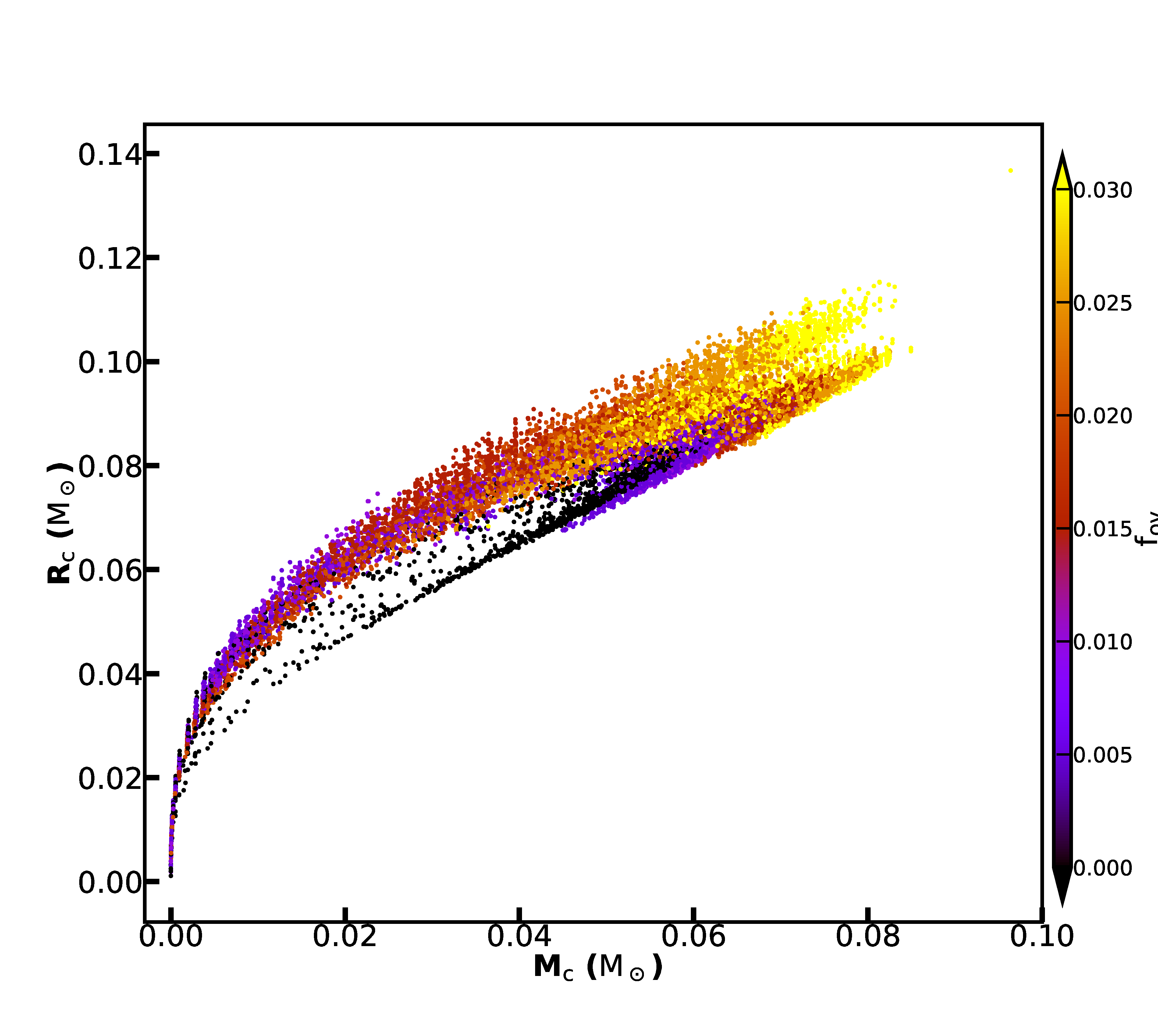}
\caption{Scatter plots showing core radius ($R_{\rm c}$) vs.~core mass ($M_{\rm c}$) for best-fit models with convective cores obtained using grid B (Run I). The top left panel is colour-coded according to the model mass ($M$), the top right panel according to the initial metal mass fraction ($Z$), and the bottom panel according to the overshoot parameter ($f_{\rm ov}$).
}
\label{variations}
\end{figure*}

The model properties that influence the onset of the CNO cycle (and associated convective core) include the adopted physics, metallicity, and mass. As mentioned in Sect.~\ref{model_grids}, both grids contain the same physics apart from the metallicity mixture. We note that models with high metallicity have a higher chance of developing convective cores. This is because a high metallicity leads to an increase in opacity, which in turn reduces the efficiency of radiative energy transport. This ultimately results in an increase in core temperature which favours the onset of the CNO cycle. Similarly, models with a higher mass have higher core temperatures, hence higher chances of developing a convective core. The top left panel of Fig.~\ref{variations} shows that best-fit models with high mass develop large and massive convective cores. This could be explained by their high overshoot parameter values as shown in the bottom panel of Fig~\ref{variations}. Best-fit models with $f_{\rm ov} =0 $ have smaller core masses and radii. The top right panel of Fig.~\ref{variations} shows no clear trend 
regarding the contribution from the initial metal mass fraction. We note that consistent results as those in Fig.~\ref{variations} are obtained when different runs are used.

It is interesting to assess the dominant model property that facilitates the occurrence of convective cores for the best-fit models in either run.
Despite the high metallicity of best-fit models from Grid A (Run II) (see bottom left panel of Fig.~\ref{other_parameters}), the majority of these models have masses $\lesssim$ 1.1 M$_\odot$ (see right panel of Fig.~\ref{mass_parameter}) resulting into 46\% of models with convective cores. Grid B (Run II) contains most of the best-fit models with low metallicity but with masses  $\gtrsim$ 1.1 M$_\odot$ (see right panel of Fig.~\ref{mass_parameter}), leading to 79\% of models with convective cores. Hence, for Run II (both grids), model mass is the dominant model property responsible for the onset of the CNO cycle. 
\begin{figure*}[!h]
\minipage{0.5\textwidth}
\hspace{-0.5cm}
  \includegraphics[width=\linewidth]{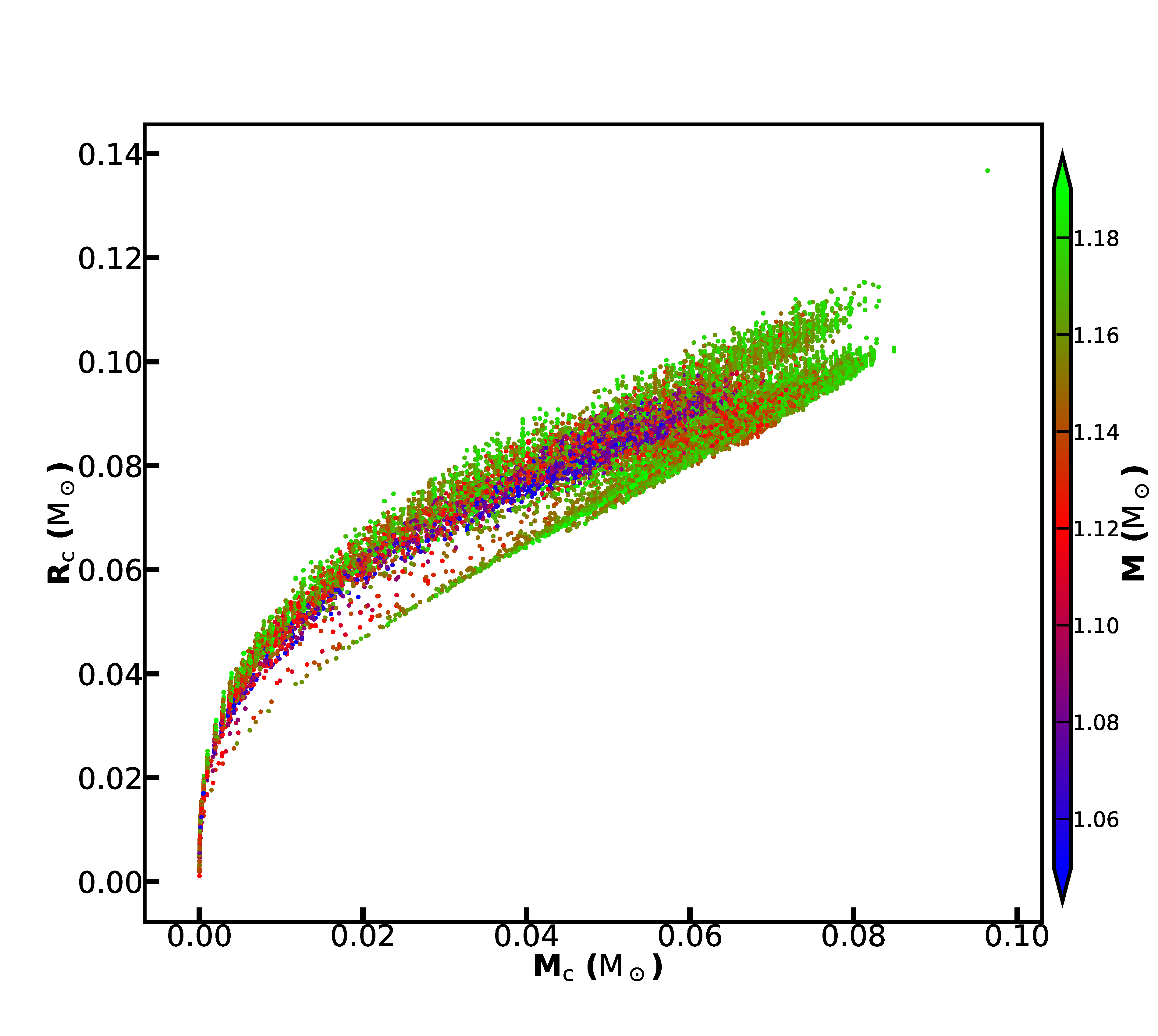}
\endminipage\hfill
\minipage{0.5\textwidth}
  \includegraphics[width=\linewidth]{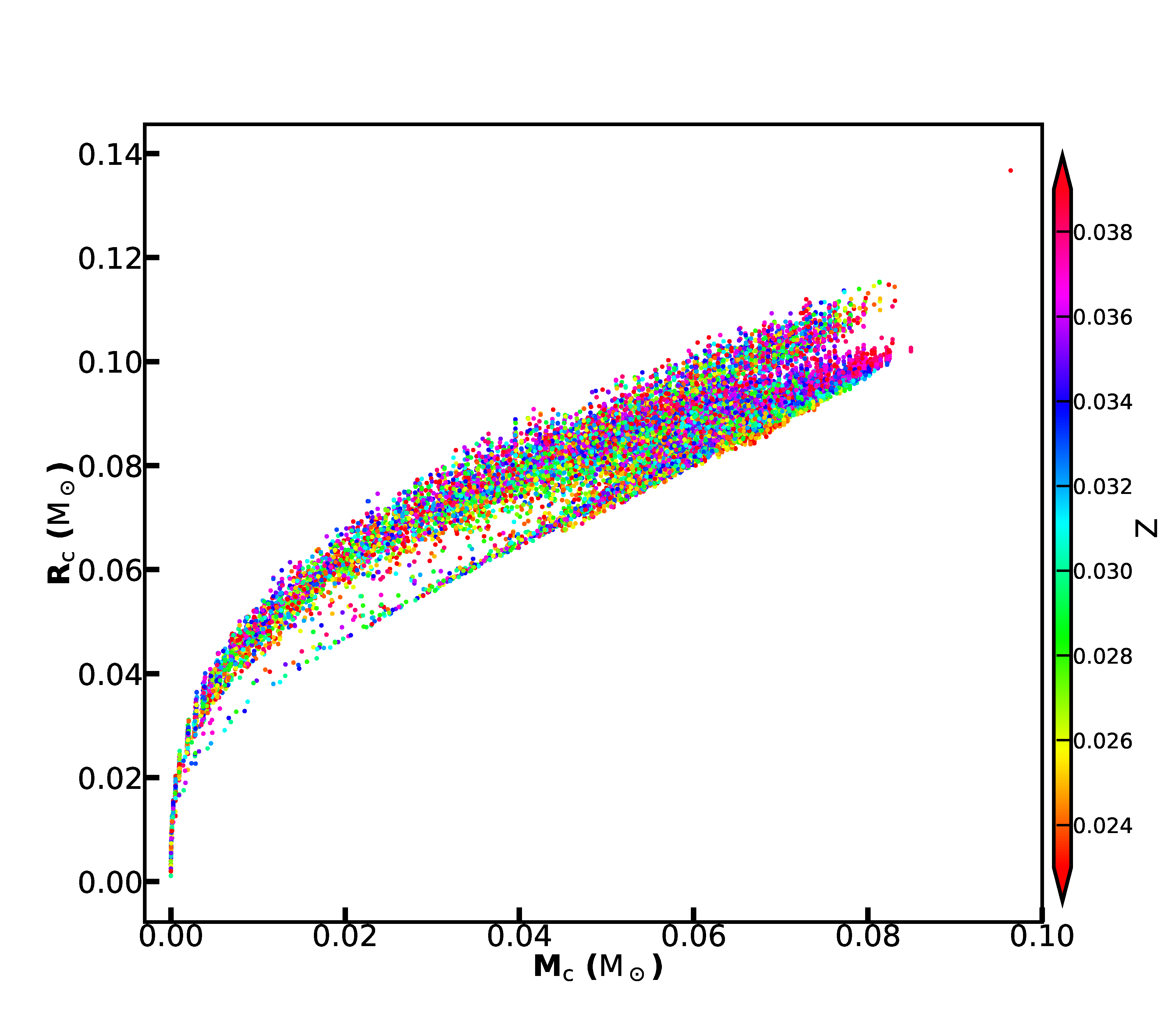}
\endminipage
\end{figure*}

For Grid A (Run I), it is challenging to determine the dominant model property that yields convective cores. This is because most of the best-fit models have masses $\gtrsim$ 1.1 M$_\odot$ and high metallicity (see left panel of Fig.~\ref{mass_parameter} and bottom left panel of Fig.~\ref{other_parameters}). However, for Grid B (Run I), the majority of best-fit models have low metallicity but masses $\gtrsim$ 1.1 M$_\odot$, with 79\% of models having convective cores. Therefore, also in this case model mass is the dominant model property contributing to the onset of the CNO cycle. Further, our results show that the mass range over which models constructed at different metallicities are expected to develop convective cores is 1.05 -- 1.15 M$_\odot$.

\section{Conclusions}
\label{conclude}

In Paper I we assessed the impact on the nature of the core of $\alpha$ Centauri A of varying the nuclear reaction rates, which showed that $\gtrsim$ 70\% of best-fit models reproducing the revised dynamical mass of \citet{Pour2016} have convective cores. In this article, we expanded on the previous work by exploring the impact of varying the metallicity mixture (and corresponding opacities). Our findings show that  $\gtrsim$ 70\% of best-fit models reproducing the revised dynamical mass have convective cores.  

In sum, the percentage of best-fit models with convective cores remains above 70\% when imposing the most up-to-date set of observational constraints. This happens irrespective of the adopted metallicity mixture and nuclear reaction rates. Therefore, we propose that $\alpha$ Centauri A be adopted in the calibration of stellar model parameters when modelling solar-like stars with small convective cores.

During the ``standard'' solar calibration process, the initial metal mass fraction ($Z$), initial helium mass fraction ($Y$), and the mixing length parameter ($\alpha_{\rm mlt}$) are varied  until a model is attained that satisfies the observed oscillation frequencies, effective temperature, metallicity, luminosity, and radius at the current solar age. The same model physics and solar calibrated parameters are then used to create grids for modelling other stars. Unlike the case of the Sun, there is no model-independent age for $\alpha$ Centauri A, but we do have a precise dynamical mass, interferometric radius, effective temperature, metallicity, and luminosity. 
In addition, we currently have ground-based seismic data, with the quality of those data expected to improve following the star's planned observations by space-based missions such as NASA's Transiting Exoplanet Survey Satellite (TESS; \citealt{Rick2015,Campante}) and ESA's PLAnetary Transits and Oscillations of stars (PLATO; \citealt{Rauer1}). This will improve the precision of observed oscillation frequencies and is also expected to increase the number of oscillation frequencies for all observable spherical degrees ($l$). This will support a more comprehensive asteroseismic analysis than the one presented here and in Paper I.

Therefore, with all these sets of observables, it will be possible to carry out a calibration procedure similar to the “standard” solar calibration routine briefly described above, without having the age among the constraints. 
It will also be feasible to provide effective constraints on some aspects of the physics, namely convection (mixing length, overshoot, surface effects), diffusion and opacities. The potential for constraining reaction rates is also a
possibility when two or all stellar components of this triple system have seismic data of high precision available.

\section*{Funding}
This work was supported by FCT - Fundação para a Ciência e a Tecnologia  through national funds and by FEDER through COMPETE2020 - Programa Operacional Competitividade e Internacionalização by these grants: UID/FIS/04434/2013 \& POCI-01-0145-FEDER-007672, PTDC/FIS-AST/30389/2017 \& POCI-01-0145-FEDER-030389 and PTDC/FIS-AST/28953/2017 \& POCI-01-0145-FEDER-028953. B.~Nsamba is supported by Funda\c{c}\~{a}o para a Ci\^{e}ncia e a Tecnologia (FCT, Portugal) under grant PD/BD/113744/2015 from PhD::SPACE, an FCT PhD program. T.~L.~Campante acknowledges support from the European Union’s Horizon 2020 research and innovation programme under the Marie Sk\l{}odowska-Curie grant agreement No.~792848 (PULSATION) and from grant CIAAUP-12/2018-BPD. 
SGS acknowledges   support   from   FCT   through   Investigador   FCT   contract No.  IF/00028/2014/CP1215/CT0002  and  from  FEDER through COMPETE2020 (grants UID/FIS/04434/2013 \& PTDC/FIS-AST/7073/2014 \& POCI-01-0145 FEDER-016880). Based on data obtained from the ESO Science Archive Facility under request number SAF Alpha Cen A 86436. T

\section*{Acknowledgements}
We acknowledge the referees for the helpful and constructive remarks.
B.~Nsamba would like to thank the MESA community for advice on how to handle specific issues regarding the fine-tuning of the MESA code so as to produce sensible results. B.~Nsamba would also like to thank the members of the asteroseismology group at the Instituto de Astrof\'{\i}sica e Ci\^{e}ncias do Espa\c{c}o for their feedback, which added value to this work.



\bibliographystyle{frontiersinSCNS_ENG_HUMS} 
\bibliography{test}

\begin{thebibliography}{52}
\providecommand{\natexlab}[1]{#1}
\expandafter\ifx\csname urlstyle\endcsname\relax
  \providecommand{\doi}[1]{doi:\discretionary{}{}{}#1}\else
  \providecommand{\doi}{doi:\discretionary{}{}{}\begingroup
  \urlstyle{rm}\Url}\fi
\providecommand{\selectlanguage}[1]{\relax}
\providecommand{\bibAnnoteFile}[1]{%
  \IfFileExists{#1}{\begin{quotation}\noindent\textsc{Key:} #1\\
  \textsc{Annotation:}\ \input{#1}\end{quotation}}{}}
\providecommand{\bibAnnote}[2]{%
  \begin{quotation}\noindent\textsc{Key:} #1\\
  \textsc{Annotation:}\ #2\end{quotation}}

\bibitem[{Aerts et~al.(2010)Aerts, Christensen-Dalsgaard, and
  Kurtz}]{Aerts2010}
Aerts, C., Christensen-Dalsgaard, J., and Kurtz, D. (2010).
\newblock \emph{Asteroseismology}.
\newblock Astronomy and Astrophysics Library (Springer Netherlands)
\bibAnnoteFile{Aerts2010}

\bibitem[{{Asplund} et~al.(2005){Asplund}, {Grevesse}, and
  {Sauval}}]{Asplund2005}
{Asplund}, M., {Grevesse}, N., and {Sauval}, A.~J. (2005).
\newblock {The Solar Chemical Composition}.
\newblock In \emph{Cosmic Abundances as Records of Stellar Evolution and
  Nucleosynthesis}, eds. T.~G. {Barnes}, III and F.~N. {Bash}. vol. 336 of
  \emph{Astronomical Society of the Pacific Conference Series}, 25
\bibAnnoteFile{Asplund2005}

\bibitem[{{Asplund} et~al.(2009){Asplund}, {Grevesse}, {Sauval}, and
  {Scott}}]{Asplund}
{Asplund}, M., {Grevesse}, N., {Sauval}, A.~J., and {Scott}, P. (2009).
\newblock {The Chemical Composition of the Sun}.
\newblock \emph{Annual Review of Astron and Astrophys} 47, 481--522.
\newblock \doi{10.1146/annurev.astro.46.060407.145222}
\bibAnnoteFile{Asplund}

\bibitem[{{Ball} and {Gizon}(2014)}]{Ball2014}
{Ball}, W.~H. and {Gizon}, L. (2014).
\newblock {A new correction of stellar oscillation frequencies for near-surface
  effects}.
\newblock \emph{A\&A} 568, A123.
\newblock \doi{10.1051/0004-6361/201424325}
\bibAnnoteFile{Ball2014}

\bibitem[{{Bazot} et~al.(2007){Bazot}, {Bouchy} et~al.}]{Bazot2007}
{Bazot}, M., {Bouchy}, F., et~al. (2007).
\newblock {Asteroseismology of {$\alpha$} Centauri A. Evidence of rotational
  splitting}.
\newblock \emph{A\&A} 470, 295--302.
\newblock \doi{10.1051/0004-6361:20065694}
\bibAnnoteFile{Bazot2007}

\bibitem[{{Bazot} et~al.(2012){Bazot}, {Bourguignon}, and
  {Christensen-Dalsgaard}}]{2012Bazot}
{Bazot}, M., {Bourguignon}, S., and {Christensen-Dalsgaard}, J. (2012).
\newblock {A Bayesian approach to the modelling of {$\alpha$} Cen A}.
\newblock \emph{MNRAS} 427, 1847--1866.
\newblock \doi{10.1111/j.1365-2966.2012.21818.x}
\bibAnnoteFile{2012Bazot}

\bibitem[{{Bazot} et~al.(2016){Bazot}, {Christensen-Dalsgaard}, {Gizon}, and
  {Benomar}}]{2016Bazot}
{Bazot}, M., {Christensen-Dalsgaard}, J., {Gizon}, L., and {Benomar}, O.
  (2016).
\newblock {On the uncertain nature of the core of {$\alpha$} Cen A}.
\newblock \emph{MNRAS} 460, 1254--1269.
\newblock \doi{10.1093/mnras/stw921}
\bibAnnoteFile{2016Bazot}

\bibitem[{{Bedding} et~al.(2005){Bedding}, {Kjeldsen}, {Bouchy}, {Bruntt},
  {Butler}, {Buzasi} et~al.}]{Bed}
{Bedding}, T.~R., {Kjeldsen}, H., {Bouchy}, F., {Bruntt}, H., {Butler}, R.~P.,
  {Buzasi}, D.~L., et~al. (2005).
\newblock {The non-detection of oscillations in Procyon by MOST: Is it really a
  surprise?}
\newblock \emph{A\&A} 432, L43--L48.
\newblock \doi{10.1051/0004-6361:200500019}
\bibAnnoteFile{Bed}

\bibitem[{{B{\"o}hm-Vitense}(1958)}]{Vitense}
{B{\"o}hm-Vitense}, E. (1958).
\newblock {{\"U}ber die Wasserstoffkonvektionszone in Sternen verschiedener
  Effektivtemperaturen und Leuchtkr{\"a}fte. Mit 5 Textabbildungen}.
\newblock \emph{Zeit. Astrophys.} 46, 108
\bibAnnoteFile{Vitense}

\bibitem[{{Bouchy} and {Carrier}(2002)}]{Bouchy}
{Bouchy}, F. and {Carrier}, F. (2002).
\newblock {The acoustic spectrum of alpha Cen A}.
\newblock \emph{A\&A} 390, 205--212.
\newblock \doi{10.1051/0004-6361:20020706}
\bibAnnoteFile{Bouchy}

\bibitem[{{Campante} et~al.(2016){Campante}, {Schofield}, {Kuszlewicz},
  {Bouma}, {Chaplin}, {Huber} et~al.}]{Campante}
{Campante}, T.~L., {Schofield}, M., {Kuszlewicz}, J.~S., {Bouma}, L.,
  {Chaplin}, W.~J., {Huber}, D., et~al. (2016).
\newblock {The Asteroseismic Potential of TESS: Exoplanet-host Stars}.
\newblock \emph{ApJ} 830, 138.
\newblock \doi{10.3847/0004-637X/830/2/138}
\bibAnnoteFile{Campante}

\bibitem[{{Compton} et~al.(2018){Compton}, {Bedding}, {Ball}, {Stello},
  {Huber}, {White} et~al.}]{Compton2018}
{Compton}, D.~L., {Bedding}, T.~R., {Ball}, W.~H., {Stello}, D., {Huber}, D.,
  {White}, T.~R., et~al. (2018).
\newblock {Surface correction of main-sequence solar-like oscillators with the
  Kepler LEGACY sample}.
\newblock \emph{MNRAS} 479, 4416--4431.
\newblock \doi{10.1093/mnras/sty1632}
\bibAnnoteFile{Compton2018}

\bibitem[{{Cyburt} et~al.(2003)}]{Cyburt2003}
{Cyburt}, R.~H. et~al. (2003).
\newblock {Primordial nucleosynthesis in light of WMAP}.
\newblock \emph{Physics Letters B} 567, 227--234.
\newblock \doi{10.1016/j.physletb.2003.06.026}
\bibAnnoteFile{Cyburt2003}

\bibitem[{Cyburt et~al.(2010)}]{Cyburt}
Cyburt, R.~H. et~al. (2010).
\newblock The jina reaclib database: Its recent updates and impact on type-i
  x-ray bursts.
\newblock \emph{ApJS} 189, 240
\bibAnnoteFile{Cyburt}

\bibitem[{{de Meulenaer} et~al.(2010)}]{Meulen2010}
{de Meulenaer}, P. et~al. (2010).
\newblock {Core properties of {$\alpha$} Centauri A using asteroseismology}.
\newblock \emph{A\&A} 523, A54.
\newblock \doi{10.1051/0004-6361/201014966}
\bibAnnoteFile{Meulen2010}

\bibitem[{{Delahaye} and {Pinsonneault}(2006)}]{Delahaye2006}
{Delahaye}, F. and {Pinsonneault}, M.~H. (2006).
\newblock {The Solar Heavy-Element Abundances. I. Constraints from Stellar
  Interiors}.
\newblock \emph{ApJ} 649, 529--540.
\newblock \doi{10.1086/505260}
\bibAnnoteFile{Delahaye2006}

\bibitem[{{Ferguson} et~al.(2005){Ferguson}, {Alexander} et~al.}]{Ferguson}
{Ferguson}, J.~W., {Alexander}, D.~R., et~al. (2005).
\newblock {Low-Temperature Opacities}.
\newblock \emph{ApJ} 623, 585--596.
\newblock \doi{10.1086/428642}
\bibAnnoteFile{Ferguson}

\bibitem[{{Grevesse} et~al.(1998)}]{Grevesse}
{Grevesse}, N. et~al. (1998).
\newblock {Standard Solar Composition}.
\newblock \emph{Space Science Reviews} 85, 161--174.
\newblock \doi{10.1023/A:1005161325181}
\bibAnnoteFile{Grevesse}

\bibitem[{{Herwig}(2000)}]{Herwig}
{Herwig}, F. (2000).
\newblock {The evolution of AGB stars with convective overshoot}.
\newblock \emph{A\&A} 360, 952--968
\bibAnnoteFile{Herwig}

\bibitem[{{Iglesias} and {Rogers}(1996)}]{Iglesias}
{Iglesias}, C.~A. and {Rogers}, F.~J. (1996).
\newblock {Updated Opal Opacities}.
\newblock \emph{ApJ} 464, 943.
\newblock \doi{10.1086/177381}
\bibAnnoteFile{Iglesias}

\bibitem[{{Imbriani} et~al.(2005){Imbriani}, {Costantini} et~al.}]{Imbriani}
{Imbriani}, G., {Costantini}, H., et~al. (2005).
\newblock {S-factor of $^{14}$N(p,{$\gamma$})$^{15}$O at astrophysical
  energies$^{⋆}$}.
\newblock \emph{EPJ A} 25, 455--466.
\newblock \doi{10.1140/epja/i2005-10138-7}
\bibAnnoteFile{Imbriani}

\bibitem[{{J{\o}rgensen} et~al.(2019){J{\o}rgensen}, {Weiss}, {Angelou}, and
  {Silva Aguirre}}]{gensen2019}
{J{\o}rgensen}, A.~C.~S., {Weiss}, A., {Angelou}, G., and {Silva Aguirre}, V.
  (2019).
\newblock {Mending the structural surface effect of 1D stellar structure models
  with non-solar metallicities based on interpolated 3D envelopes}.
\newblock \emph{MNRAS} 484, 5551--5567.
\newblock \doi{10.1093/mnras/stz337}
\bibAnnoteFile{gensen2019}

\bibitem[{{Joyce} and {Chaboyer}(2018)}]{2018Joyce}
{Joyce}, M. and {Chaboyer}, B. (2018).
\newblock {Classically and Asteroseismically Constrained 1D Stellar Evolution
  Models of {$\alpha$} Centauri A and B Using Empirical Mixing Length
  Calibrations}.
\newblock \emph{ApJ} 864, 99.
\newblock \doi{10.3847/1538-4357/aad464}
\bibAnnoteFile{2018Joyce}

\bibitem[{{Kervella} et~al.(2017){Kervella}, {Bigot} et~al.}]{Kervella2017}
{Kervella}, P., {Bigot}, L., et~al. (2017).
\newblock {The radii and limb darkenings of {$\alpha$} Centauri A and B .
  Interferometric measurements with VLTI/PIONIER}.
\newblock \emph{A\&A} 597, A137.
\newblock \doi{10.1051/0004-6361/201629505}
\bibAnnoteFile{Kervella2017}

\bibitem[{{Kervella} et~al.(2016){Kervella}, {Mignard} et~al.}]{Kervella2016}
{Kervella}, P., {Mignard}, F., et~al. (2016).
\newblock {Close stellar conjunctions of {$\alpha$} Centauri A and B until 2050
  . An m$_{K}$ = 7.8 star may enter the Einstein ring of {$\alpha$} Cen A in
  2028}.
\newblock \emph{A\&A} 594, A107.
\newblock \doi{10.1051/0004-6361/201629201}
\bibAnnoteFile{Kervella2016}

\bibitem[{{Kim}(1999)}]{1999Kim}
{Kim}, Y.-C. (1999).
\newblock {Standard Stellar Models; alpha Cen A and B}.
\newblock \emph{Journal of Korean Astronomical Society} 32, 119--126
\bibAnnoteFile{1999Kim}

\bibitem[{{Kunz} et~al.(2002){Kunz}, {Fey} et~al.}]{Kunz}
{Kunz}, R., {Fey}, M., et~al. (2002).
\newblock {Astrophysical Reaction Rate of $^{12}$C({$\alpha$},
  {$\gamma$})$^{16}$O}.
\newblock \emph{ApJ} 567, 643--650.
\newblock \doi{10.1086/338384}
\bibAnnoteFile{Kunz}

\bibitem[{{Lebreton} and {Goupil}(2014)}]{Lebreton}
{Lebreton}, Y. and {Goupil}, M.~J. (2014).
\newblock {Asteroseismology for ``{\`a} la carte'' stellar age-dating and
  weighing. Age and mass of the CoRoT exoplanet host HD 52265}.
\newblock \emph{A\&A} 569, A21.
\newblock \doi{10.1051/0004-6361/201423797}
\bibAnnoteFile{Lebreton}

\bibitem[{{Lund} and {Reese}(2018)}]{Reese}
{Lund}, M.~N. and {Reese}, D.~R. (2018).
\newblock {Tutorial: Asteroseismic Stellar Modelling with AIMS}.
\newblock \emph{ASSSP} 49, 149.
\newblock \doi{10.1007/978-3-319-59315-9_8}
\bibAnnoteFile{Reese}

\bibitem[{{Mathur} et~al.(2012){Mathur}, {Metcalfe}, {Woitaszek}, {Bruntt},
  {Verner}, {Christensen-Dalsgaard} et~al.}]{Mathur}
{Mathur}, S., {Metcalfe}, T.~S., {Woitaszek}, M., {Bruntt}, H., {Verner},
  G.~A., {Christensen-Dalsgaard}, J., et~al. (2012).
\newblock {A Uniform Asteroseismic Analysis of 22 Solar-type Stars Observed by
  Kepler}.
\newblock \emph{ApJ} 749, 152.
\newblock \doi{10.1088/0004-637X/749/2/152}
\bibAnnoteFile{Mathur}

\bibitem[{{Metcalfe} et~al.(2012){Metcalfe}, {Chaplin}, {Appourchaux},
  {Garc{\'{\i}}a}, {Basu}, {Brand{\~a}o} et~al.}]{Still}
{Metcalfe}, T.~S., {Chaplin}, W.~J., {Appourchaux}, T., {Garc{\'{\i}}a}, R.~A.,
  {Basu}, S., {Brand{\~a}o}, I., et~al. (2012).
\newblock {Asteroseismology of the Solar Analogs 16 Cyg A and B from Kepler
  Observations}.
\newblock \emph{APJL} 748, L10.
\newblock \doi{10.1088/2041-8205/748/1/L10}
\bibAnnoteFile{Still}

\bibitem[{{Metcalfe} et~al.(2014){Metcalfe}, {Creevey}, {Do{\u g}an}, {Mathur},
  {Xu}, {Bedding} et~al.}]{Metcalfe}
{Metcalfe}, T.~S., {Creevey}, O.~L., {Do{\u g}an}, G., {Mathur}, S., {Xu}, H.,
  {Bedding}, T.~R., et~al. (2014).
\newblock {Properties of 42 Solar-type Kepler Targets from the Asteroseismic
  Modeling Portal}.
\newblock \emph{ApJS} 214, 27.
\newblock \doi{10.1088/0067-0049/214/2/27}
\bibAnnoteFile{Metcalfe}

\bibitem[{{Miglio} and {Montalb{\'a}n}(2005)}]{Monta}
{Miglio}, A. and {Montalb{\'a}n}, J. (2005).
\newblock {Constraining fundamental stellar parameters using seismology.
  Application to {$\alpha$} Centauri AB}.
\newblock \emph{A\&A} 441, 615--629.
\newblock \doi{10.1051/0004-6361:20052988}
\bibAnnoteFile{Monta}

\bibitem[{{Nsamba} et~al.(2018{\natexlab{a}}){Nsamba}, {Monteiro}, {Campante},
  {Cunha}, and {Sousa}}]{201Nsamba}
{Nsamba}, B., {Monteiro}, M.~J.~P.~F.~G., {Campante}, T.~L., {Cunha}, M.~S.,
  and {Sousa}, S.~G. (2018{\natexlab{a}}).
\newblock {{$\alpha$} Centauri A as a potential stellar model calibrator:
  establishing the nature of its core}.
\newblock \emph{MNRAS} 479, L55--L59.
\newblock \doi{10.1093/mnrasl/sly092}
\bibAnnoteFile{201Nsamba}

\bibitem[{{Nsamba} et~al.(2018{\natexlab{b}})}]{2018Nsamba}
{Nsamba}, B. et~al. (2018{\natexlab{b}}).
\newblock {Asteroseismic modelling of solar-type stars: internal systematics
  from input physics and surface correction methods}.
\newblock \emph{MNRAS} arXiv:1804.04935.
\newblock \doi{10.1093/mnras/sty948}
\bibAnnoteFile{2018Nsamba}

\bibitem[{{Paxton} et~al.(2011){Paxton}, {Bildsten}, {Dotter}, {Herwig},
  {Lesaffre}, and {Timmes}}]{Pax1}
{Paxton}, B., {Bildsten}, L., {Dotter}, A., {Herwig}, F., {Lesaffre}, P., and
  {Timmes}, F. (2011).
\newblock {Modules for Experiments in Stellar Astrophysics (MESA)}.
\newblock \emph{ApJS} 192, 3.
\newblock \doi{10.1088/0067-0049/192/1/3}
\bibAnnoteFile{Pax1}

\bibitem[{{Paxton} et~al.(2013){Paxton}, {Cantiello}, {Arras}, {Bildsten},
  {Brown}, {Dotter} et~al.}]{Pax2}
{Paxton}, B., {Cantiello}, M., {Arras}, P., {Bildsten}, L., {Brown}, E.~F.,
  {Dotter}, A., et~al. (2013).
\newblock {Modules for Experiments in Stellar Astrophysics (MESA): Planets,
  Oscillations, Rotation, and Massive Stars}.
\newblock \emph{ApJS} 208, 4
\bibAnnoteFile{Pax2}

\bibitem[{{Paxton} et~al.(2015){Paxton}, {Marchant}, {Schwab}, {Bauer},
  {Bildsten}, {Cantiello} et~al.}]{Pax3}
{Paxton}, B., {Marchant}, P., {Schwab}, J., {Bauer}, E.~B., {Bildsten}, L.,
  {Cantiello}, M., et~al. (2015).
\newblock {Modules for Experiments in Stellar Astrophysics (MESA): Binaries,
  Pulsations, and Explosions}.
\newblock \emph{ApJS} 220, 15.
\newblock \doi{10.1088/0067-0049/220/1/15}
\bibAnnoteFile{Pax3}

\bibitem[{{Paxton} et~al.(2018){Paxton}, {Schwab}, {Bauer}, {Bildsten},
  {Blinnikov}, {Duffell} et~al.}]{Pax2018}
{Paxton}, B., {Schwab}, J., {Bauer}, E.~B., {Bildsten}, L., {Blinnikov}, S.,
  {Duffell}, P., et~al. (2018).
\newblock {Modules for Experiments in Stellar Astrophysics (MESA): Convective
  Boundaries, Element Diffusion, and Massive Star Explosions}.
\newblock \emph{ApJS} 234, 34.
\newblock \doi{10.3847/1538-4365/aaa5a8}
\bibAnnoteFile{Pax2018}

\bibitem[{{Pinheiro} and {Fernandes}(2013)}]{Pinheiro}
{Pinheiro}, F.~J.~G. and {Fernandes}, J. (2013).
\newblock {On the (non-)universality of the mixing length parameter}.
\newblock \emph{MNRAS} 433, 2893--2899.
\newblock \doi{10.1093/mnras/stt910}
\bibAnnoteFile{Pinheiro}

\bibitem[{{Pourbaix} and {Boffin}(2016)}]{Pour2016}
{Pourbaix}, D. and {Boffin}, H.~M.~J. (2016).
\newblock {Parallax and masses of {$\alpha$} Centauri revisited}.
\newblock \emph{A\&A} 586, A90.
\newblock \doi{10.1051/0004-6361/201527859}
\bibAnnoteFile{Pour2016}

\bibitem[{{Pourbaix} et~al.(2002){Pourbaix}, {Nidever} et~al.}]{Pour2002}
{Pourbaix}, D., {Nidever}, D., et~al. (2002).
\newblock {Constraining the difference in convective blueshift between the
  components of alpha Centauri with precise radial velocities}.
\newblock \emph{A\&A} 386, 280--285.
\newblock \doi{10.1051/0004-6361:20020287}
\bibAnnoteFile{Pour2002}

\bibitem[{{Rauer} et~al.(2014){Rauer}, {Catala}, {Aerts}, {Appourchaux},
  {Benz}, {Brandeker} et~al.}]{Rauer1}
{Rauer}, H., {Catala}, C., {Aerts}, C., {Appourchaux}, T., {Benz}, W.,
  {Brandeker}, A., et~al. (2014).
\newblock {The PLATO 2.0 mission}.
\newblock \emph{Experimental Astronomy} 38, 249--330.
\newblock \doi{10.1007/s10686-014-9383-4}
\bibAnnoteFile{Rauer1}

\bibitem[{{Rendle} et~al.(2019){Rendle}, {Buldgen}, {Miglio}, {Reese}, {Noels},
  {Davies} et~al.}]{Rend2019}
{Rendle}, B.~M., {Buldgen}, G., {Miglio}, A., {Reese}, D., {Noels}, A.,
  {Davies}, G.~R., et~al. (2019).
\newblock {AIMS - a new tool for stellar parameter determinations using
  asteroseismic constraints}.
\newblock \emph{MNRAS} 484, 771--786.
\newblock \doi{10.1093/mnras/stz031}
\bibAnnoteFile{Rend2019}

\bibitem[{{Ricker} et~al.(2015){Ricker}, {Winn}, {Vanderspek}, {Latham},
  {Bakos}, {Bean} et~al.}]{Rick2015}
{Ricker}, G.~R., {Winn}, J.~N., {Vanderspek}, R., {Latham}, D.~W., {Bakos},
  G.~{\'A}., {Bean}, J.~L., et~al. (2015).
\newblock {Transiting Exoplanet Survey Satellite (TESS)}.
\newblock \emph{Journal of Astronomical Telescopes, Instruments, and Systems}
  1, 014003.
\newblock \doi{10.1117/1.JATIS.1.1.014003}
\bibAnnoteFile{Rick2015}

\bibitem[{{Serenelli} and {Basu}(2010)}]{Serenelli}
{Serenelli}, A.~M. and {Basu}, S. (2010).
\newblock {Determining the Initial Helium Abundance of the Sun}.
\newblock \emph{ApJ} 719, 865--872.
\newblock \doi{10.1088/0004-637X/719/1/865}
\bibAnnoteFile{Serenelli}

\bibitem[{{Silva Aguirre} et~al.(2015){Silva Aguirre}, {Davies}, {Basu},
  {Christensen-Dalsgaard}, {Creevey}, {Metcalfe} et~al.}]{Aguirre}
{Silva Aguirre}, V., {Davies}, G.~R., {Basu}, S., {Christensen-Dalsgaard}, J.,
  {Creevey}, O., {Metcalfe}, T.~S., et~al. (2015).
\newblock {Ages and fundamental properties of Kepler exoplanet host stars from
  asteroseismology}.
\newblock \emph{MNRAS} 452, 2127--2148.
\newblock \doi{10.1093/mnras/stv1388}
\bibAnnoteFile{Aguirre}

\bibitem[{{Silva Aguirre} et~al.(2017){Silva Aguirre}, {Lund}, {Antia}, {Ball},
  {Basu}, {Christensen-Dalsgaard} et~al.}]{Aguirre1}
{Silva Aguirre}, V., {Lund}, M.~N., {Antia}, H.~M., {Ball}, W.~H., {Basu}, S.,
  {Christensen-Dalsgaard}, J., et~al. (2017).
\newblock {Standing on the Shoulders of Dwarfs: the Kepler Asteroseismic LEGACY
  Sample. II.Radii, Masses, and Ages}.
\newblock \emph{ApJ} 835, 173.
\newblock \doi{10.3847/1538-4357/835/2/173}
\bibAnnoteFile{Aguirre1}

\bibitem[{{S{\"o}derhjelm}(1999)}]{derhjelm}
{S{\"o}derhjelm}, S. (1999).
\newblock {Visual binary orbits and masses POST HIPPARCOS}.
\newblock \emph{A\&A} 341, 121--140
\bibAnnoteFile{derhjelm}

\bibitem[{{Thoul} et~al.(1994)}]{Thoul}
{Thoul}, A.~A. et~al. (1994).
\newblock {Element diffusion in the solar interior}.
\newblock \emph{ApJ} 421, 828--842.
\newblock \doi{10.1086/173695}
\bibAnnoteFile{Thoul}

\bibitem[{{Townsend} and {Teitler}(2013)}]{Townsend}
{Townsend}, R.~H.~D. and {Teitler}, S.~A. (2013).
\newblock {GYRE: an open-source stellar oscillation code based on a new Magnus
  Multiple Shooting scheme}.
\newblock \emph{MNRAS} 435, 3406--3418.
\newblock \doi{10.1093/mnras/stt1533}
\bibAnnoteFile{Townsend}

\bibitem[{{Y{\i}ld{\i}z}(2007)}]{Yild}
{Y{\i}ld{\i}z}, M. (2007).
\newblock {Models of {$\alpha$} Centauri A and B with and without seismic
  constraints: time dependence of the mixing-length parameter}.
\newblock \emph{MNRAS} 374, 1264--1270.
\newblock \doi{10.1111/j.1365-2966.2006.11218.x}
\bibAnnoteFile{Yild}

\end{thebibliography}

\end{document}